\documentclass[11pt,a4paper]{article}
\usepackage{mathrsfs}
\usepackage{bm}
\usepackage{amsmath,amssymb}
\usepackage[colorlinks]{hyperref}
\usepackage{graphicx}
\usepackage[top=25truemm,bottom=25truemm,left=20truemm,right=20truemm]{geometry}
\makeatletter
	
		\@addtoreset{equation}{section}
\makeatother
\title{Exact solutions of primordial gravitational waves}
\author{Tadashi Sasaki\footnote{t-sasaki@particle.sci.hokudai.ac.jp}  and  Hisao Suzuki\footnote{hsuzuki@particle.sci.hokudai.ac.jp} \\[5pt] \emph{Department of Physics, Hokkaido University, Sapporo 060-0810, Japan}}
\date{}
\begin{document}
\maketitle
\abstract{
	The future detection projects of gravitational waves are expected to have the sensitivity of detecting the primordial gravitational waves, which may be useful to get new insight into the very early universe. It is essential to analyze the evolution equation of the gravitational waves to estimate the present field strength of the primordial gravitational waves.  
	
	In this paper, we obtain analytic solutions of the gravitational wave equation in the presence of non-relativistic matters as well as the cosmological constant.  Although it is difficult to obtain the solution directly, we find that the equation for the square of the amplitudes has a simple polynomial solution.
	This quantity, which is directly related to the energy density of the gravitational waves, turns out to be useful to construct analytic solutions for the amplitudes by using Weierstrass's elliptic functions.
}

\section{Introduction} 
  The direct detection of gravitational waves \cite{LIGO} opened an era of gravitational wave astronomy.  
The ground-based experiments, such as KAGRA\cite{KAGRA}, Advanced LIGO\cite{ALIGO}, and Advanced VIRGO\cite{AVIRGO} 
are expected to observe many gravitational wave signals from binary systems of black holes and neutron stars, 
which may help us investigate various merging processes. The future space-based experiments such as eLISA\cite{eLISA}, 
DECIGO\cite{DECIGO}, and BBO\cite{BBO} are expected to have the sensitivity to investigate the primordial gravitational waves\cite{Grishchuk}, 
namely, the gravitational waves produced in the early stage of the universe. 
These primordial gravitational waves are excellent tools to investigate some cosmological phenomena such as 
1st order phase transition\cite{KKM} and the inflation\cite{TWL}.
  
As for the predictions of the gravitational waves, the production and the evolution are two essential factors of the analysis. 
However, in most cases where the transition of the era is concerned, it is difficult to get any analytic solutions of the evolution equations. 
For example, the transition from the radiation-dominant era to the matter-dominant era has been analyzed in ref.\cite{TWL}. 
It was shown by numerical calculation that the transition function can be obtained in a rather simple form. 
Later, various numerical improvements and analytical approximations have been achieved\cite{PK,Zhao,WatanabeKomatsu,SakuraiShirai}. However, no analytic solution has been found yet.
  The main reason for the difficulty is that the equation has too many singular points. 
If an ordinary second-order differential equation has three definite singular points or the confluent limit of them, 
it is well known that the solution is given by the hypergeometric (or confluent hypergeometric) function, 
where various analytic continuations are applicable. In the case of an equation with four definite singular points, which is called
Heun's equation (see for example \cite{Heun}), we cannot usually get any analytical expression, so that it is difficult 
to expect the matching beyond the region of the convergence in the power series solutions. 
There are several exceptions, however. For example, analytic solutions have been obtained for Lam{\'{e}} equation, which has four singular points. 
As we will see later, the generic evolution equation of the gravitational waves has six singular points, 
so that we do not expect to have any analytic solutions. In the case of radiation to matter transition, 
the equation is the confluent Heun's equation.
  
  The technique for solving the equations appeared in 2004\cite{Takemura1}. 
The technique is basically for Lam{\'{e}} equations\cite{BLH} and later extended to the case where the equation has more singularities\cite{Takemura2}. 
The basic idea of these papers are the followings: we solve the third order equation for the square of the solutions of the 
original equation instead of solving the equation directly.
  
In this paper, we will show that for the transition from the matter-dominant era to the cosmological-constant-dominant era,
the evolution of the square of the amplitude of gravitational waves can be expressed as a polynomial of the reciprocal of the scale factor $a$,
which is used to construct exact solutions to the gravitational wave equation in terms of Weierstrass elliptic function $\wp$.

In the next section, we will review the evolution equation in the cosmological setting.
In section 3, we will explain the formalism of solving the second order equation out of the corresponding third order equation.
We will apply the formalism to the gravitational wave equation and get exact results in the presence of matters and the cosmological constant in section 4 and 5.
The final section will be devoted to discussions.

\section{The Equation for gravitational waves}
We consider homogeneous and isotropic background, namely the Friedmann-Robertson-Walker metric of the following form:
\begin{equation}
ds^2=a^2(\eta)\left(-d\eta^2 +\frac{dr^2}{1-Kr^2}+r^2 d\Omega^2\right).
\end{equation}
Here, we use the conformal time $\eta$ as a time coordinate and $K$ represents spatial curvature.
Then, time evolution of the scale factor $a$ is determined by the Friedmann equation,
\begin{equation}
\left(\frac{a^\prime}{a^2}\right)^2=H_0^2\left(\frac{\Omega_r}{a^4}+\frac{\Omega_m}{a^3}+\frac{\Omega_K}{a^2}+\Omega_\Lambda\right), \label{eq:FE1}
\end{equation}
where $H_0$ is the present $(a=1)$ Hubble constant, $a'=da/d\eta$, and
\begin{equation}
\Omega_r=\frac{8\pi G}{3}\frac{\rho_{r,0}}{H_0^2},\qquad \Omega_m=\frac{8\pi G}{3}\frac{\rho_{m,0}}{H_0^2}, \qquad \Omega_K=-\frac{K}{H_0^2},\qquad \Omega_\Lambda=\frac{\Lambda}{3H_0^2}.
\end{equation}
Here, $G$ is the Newton constant, $\rho_{r,0}$ and $\rho_{m,0}$ denote energy density of relativistic and non-relativistic matter at present respectively, 
and $\Lambda$ is the cosmological constant.
Defining a function $g(a)$ as
\begin{equation}
g(a)=\Omega_r + \Omega_m a +\Omega_K a^2 +\Omega_\Lambda a^4,
\end{equation}
we can write the equation (\ref{eq:FE1}) as
\begin{equation}
\frac{da}{d\eta}=H_0 g^{\frac{1}{2}}(a). \label{eq;FE2}
\end{equation}

The gravitational wave equation (tensor mode) is given by
\begin{equation}
h^{\prime\prime}_{ij}+2\left(\frac{a^\prime}{a}\right)h^{\prime}_{ij}+(2K-\Delta)h_{ij}=0.
\end{equation}
We decompose $h_{ij}$ as
\begin{equation}
h_{ij}(\eta,\bm{x})=\sum_{A}\sum_{k}e^A_{ij}(\bm{k})h_A(\bm{k})\chi(\eta,\bm{k})\Phi_k(\bm{x}),
\end{equation}
where $\Phi_k(\bm{x})$ is a solution for $(\Delta-2K) \Phi_k=-k^2\Phi_k$, $A=+,\times$ specifies two independent polarization states, 
$h_A(\bm{k})$ represents the initial condition of the polarization, 
and $e_{ij}^A(k)$ are the spin-2 polarization tensor satisfying the normalization condition $\sum_{ij}e_{ij}^A(e_{ij}^{B})^*=2\delta^{AB}$.
Then, the equation for $\chi(\eta,\bm{k})$  is given by
\begin{equation}
\chi^{\prime\prime}+2\left(\frac{a^\prime}{a}\right)\chi^\prime+k^2\chi=0. \label{eq:chi}
\end{equation}
To observe the generic structure of the equation, it is convenient to change the variable $\eta$ to $a$. Using the equation (\ref{eq;FE2}), we have
\begin{equation}
\left(\frac{d}{da}\right)^2\chi+\left[\frac{1}{2} \frac{g^\prime(a)}{g(a)}+\frac{2}{a}\right] \frac{d}{da}\chi+\frac{(k/H_0)^2}{g(a)}\chi=0. \label{eqchiina}
\end{equation}
Since $g(a)$ is a quartic polynomial, this equation has singular points at $g(a)=0, a=0$, and $a=\infty$, in total six singular points.  

If the equation has three definite singular points, 
the solution can be written by hypergeometric functions and we can use various analytic continuation. 
If the equation has four definite singular points, the equation is known as Heun's differential equation 
where we cannot obtain any good analytic continuation because these are essentially power series solution although many attempts has been done. 
When we include the radiation, matter and the curvature, i.e. $g=\Omega_r+\Omega_m a+\Omega_Ka^2$, 
the above equation reduces to the Heun's equation and further to the confluent Heun's equation in the limit $\Omega_K\rightarrow 0$. 
This is the reason why we do not have any analytic solution for the transition from radiation dominant to matter dominant era.

 We are going to deal with matter + cosmological constant dominant universe, which covers most of the cosmological time. 
In this case, the equation (\ref{eqchiina}) still has five singular points. Normally we do not expect the equation has any analytic solution.

\section{The third order equation associated with the second order equation}
We will follow the argument of ref.\cite{BLH}. Let us consider a generic second order differential equation of the following form:
\begin{equation}
	\chi''(x)+p(x)\chi'(x)+q(x)\chi(x)=0, \label{eq;secondorder}
\end{equation}
where $p(x)$ and $q(x)$ are known functions.
We consider a product of solutions $y(x)=\chi_1(x)\chi_2(x),$ where $\chi_1(x),\chi_2(x)$ are any solutions of the equation (\ref{eq;secondorder}).
Then the derivative can be written as
\begin{equation}
	y^\prime = \chi_1^\prime(x)\chi_2(x)+\chi_1(x)\chi_2^\prime(x). \label{yderiv}
\end{equation}
Differentiating the both sides and using the equation (\ref{eq;secondorder}), we get
\begin{equation}
	y^{\prime\prime}=-py^\prime-2qy+2\chi_1^\prime\chi_2^\prime,
\end{equation}
from which we have
\begin{equation}
	\chi_1^\prime\chi_2^\prime=\frac{1}{2}(y^{\prime\prime}+py^\prime+2qy).\label{eq:primeprime}
\end{equation}
Then it is straightforward to obtain the equation for $y(x)$ as follows:
\begin{equation}
	y^{\prime\prime\prime}+3py^{\prime\prime}+(p^\prime+4q+2p^2)y^\prime+(2q^\prime+4pq)y=0.\label{eq:thirdorder}
\end{equation}

In the second order equation of the form (\ref{eq;secondorder}),  it is well known that we can construct a constant from the Wronskian of $\chi_1(x)$ and $\chi_2(x)$ 
as
\begin{equation}
	C(\chi_1,\chi_2)=\exp\left(\int p(x)dx\right)(\chi_1\chi_2^\prime-\chi_1^\prime \chi_2),
\end{equation}
and $C(\chi_1,\chi_2)$ is nonzero if and only if the two solutions $\chi_1$ and $\chi_2$ are linearly independent.
We expect that we can get a constant from two independent solution of the equation (\ref{eq:thirdorder}).
For two solutions $y_1=\chi_1\chi_2, y_2=\chi_3\chi_4$, we can define the following constant $L$ out of the Wronskian,
\begin{equation}
	L(y_1,y_2)=C(\chi_1,\chi_3)C(\chi_2,\chi_4)+C(\chi_1,\chi_4)C(\chi_2,\chi_3)\label{eq:wronskian1}.
\end{equation}
It is easy to see that this constant (\ref{eq:wronskian1}) can be written in terms of $y_1$ and $y_2$ as 
\begin{equation}
	L(y_1,y_2)=\exp\left(2\int p(x)dx\right)[y_1(y^{\prime\prime}_2+py^\prime_2+2qy_2)+y_2(y^{\prime\prime}_1+py^\prime_1+2qy_1)-y^\prime_1y^\prime_2].
\end{equation}
This result can be used to find the $C(\chi_1,\chi_2)$ directly from the solutions of  the third order equation (\ref{eq:thirdorder}).
Actually, the square of the constant $C(\chi_1,\chi_2)$ can be obtained by equating $y_2=y_1\equiv y=\chi_1\chi_2$;
\begin{equation}
	C^2(\chi_1,\chi_2)=-\exp\left(2\int p(x)dx\right)[2yy^{\prime\prime}+2pyy^\prime+4qy^2-(y^\prime)^2]. \label{eq:csquare}
\end{equation}

Given a solution of the third order equation (\ref{eq:thirdorder}), which is assumed to have a nonzero constant 
(\ref{eq:csquare}),\footnote{When $C=0$, $y$ must be the square of some solution, i.e. $y=\chi^2$.} 
we can construct two independent solutions of (\ref{eq;secondorder}) as follows.
From (\ref{eq:csquare}), we can obtain the value of $C(\chi_1,\chi_2)$ so that
\begin{equation}
	C \exp\left(-\int p(x)dx\right)=\chi_1\chi_2^\prime-\chi_1^\prime \chi_2.
\end{equation}
Dividing this by $y=\chi_1\chi_2$, we get
\begin{equation}
	\exp\left(-\int p(x)dx\right)\frac{C}{y}=\frac{\chi^\prime_2}{\chi_2}-\frac{\chi^\prime_1}{\chi_1}. \label{eq1}
\end{equation}
On the other hand, equation (\ref{yderiv}) gives
\begin{equation}
	\frac{y^\prime}{y}= \frac{\chi^\prime_2}{\chi_2}+\frac{\chi^\prime_1}{\chi_1}. \label{eq2}
\end{equation}
Combining (\ref{eq1}) and (\ref{eq2}), we get
\begin{align}
	\frac{\chi^\prime_1}{\chi_1}=\frac{1}{2}\frac{y^\prime}{y}-\frac{1}{2} \exp\left(-\int p(x)dx\right)\frac{C}{y}\nonumber, \\
	\frac{\chi^\prime_2}{\chi_2}=\frac{1}{2}\frac{y^\prime}{y}+\frac{1}{2} \exp\left(-\int p(x)dx\right)\frac{C}{y}.
\end{align}
Integration of the above leads to the following solutions,
\begin{equation}
	\chi_{1,2}=\sqrt{y}\exp\left[\pm \frac{1}{2}\int \left(\frac{C}{y}\exp\left(-\int p(x)dx\right)\right) dx\right].
\end{equation}
Since, for wave-like equations, the constant $C$ is usually pure imaginary as we will see later, we write $C=i\tilde{C}$ so that
\begin{equation}
	\chi_{\pm}=\sqrt{y}\exp\left[\pm \frac{i}{2}\int \frac{\tilde{C}}{y}\exp\left(-\int p(x)dx\right) dx\right].\label{eq:solutions}
\end{equation}
Another choice of the solution is given by
\begin{align}
	\chi_{1}=\sqrt{y}\sin\left[\frac{1}{2}\int \frac{\tilde{C}}{y}\exp\left(-\int p(x)dx\right) dx\right],\nonumber\\
	\chi_{2}=\sqrt{y}\cos\left[ \frac{1}{2}\int \frac{\tilde{C}}{y}\exp\left(-\int p(x)dx\right) dx\right].
\end{align}
If we can find non-oscillatory solution $y$ for the third order equation (\ref{eq:thirdorder}), 
the solutions (\ref{eq:solutions}) can be regarded as phase space representation of the solutions.
In our application, the construction of the solution above has a significant advantage. 
In physics, the power of the wave can be written as the time average ($x=\eta$ or $x=a$ in the cosmological setting) of the square of the field strength. 
In any form of the wave equation, the power is proportional to $y$. Therefore, the evolution of $y$ is physically significant.

\paragraph{}
We are going to give simple examples.
The first one is the equation for a harmonic oscillator
\begin{equation}
	\chi''(x)+k^2\chi(x)=0.
\end{equation}
In this case, we know the solutions very well,
\begin{equation}
	\chi_1(x)=\cos kx, \ \chi_2(x)=\sin kx.
\end{equation}
The generic solutions are arbitrary linear combinations of these two independent solutions.
The associated third order equation (\ref{eq:thirdorder}) becomes
\begin{equation}
	y^{\prime\prime\prime}+4k^2y^\prime=0,
\end{equation}
whose independent solutions are as follows:
\begin{equation}
	y_0=1,\qquad y_1=\cos 2kx,\qquad y_2=\sin 2kx.
\end{equation}
If we take a solution $y=A^2$ ($A$ : const.), then from (\ref{eq:csquare}) 
we can see, as mentioned before, the constant $C$ becomes pure imaginary,
\begin{equation}
	C=2ik A^2.
\end{equation}
The construction of the solution (\ref{eq:solutions}) tells us that
\begin{equation}
	\chi_{\pm}=Ae^{\pm ikx}.
\end{equation}
Therefore, $y=A^2=|\chi_\pm|^2$ is the square of the amplitude $\chi_\pm$ and is a constant in this case.

\begin{figure}[t]
\centering
\includegraphics[clip,width=10.0cm]{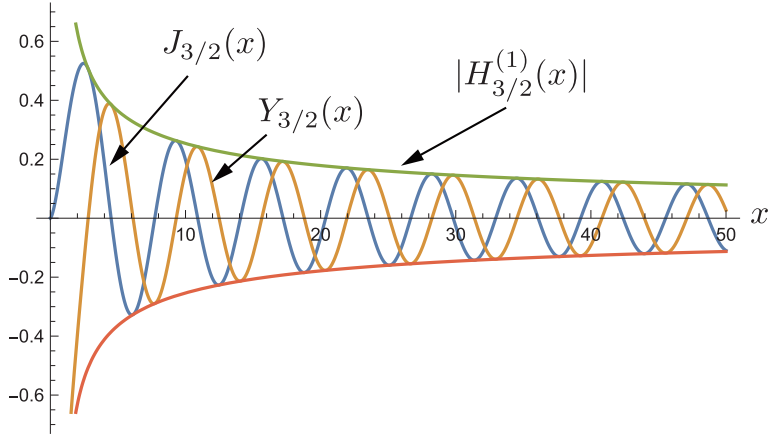}
\caption{Functions, $J_{3/2}(x),Y_{3/2}(x),\vert H_{3/2}^{(1)} (x)\vert$, the square root of $H_{3/2}^{(1)} (x)H_{3/2}^{(2)} (x)$ which is a simple polynomial representing the amplitude of the oscillations.}
\label{fig:bessel}
\end{figure}

\paragraph{}
As a next example, let us consider the Bessel's equation of order $\nu$,
\begin{equation}
	\chi^{\prime\prime}+\frac{1}{x}\chi^{\prime}+\left(1-\frac{\nu^2}{x^2}\right)\chi=0.\label{eq:besselequation}
\end{equation}
Then the third order equation (\ref{eq:thirdorder}) can be written as
\begin{equation}
	[\theta(\theta+2\nu)(\theta-2\nu)+4x^2(\theta+1)]y=0,
\end{equation}
where 
\begin{equation}
	\theta=x\frac{d}{dx}.
\end{equation}
It is easy to get three independent power series solutions as
\begin{align}
	y_{\nu}(x)&=\sum_{n=0}^\infty\frac{(\nu+1/2)_n}{(\nu+1)_n(2\nu+1)_nn!}(-x^2)^nx^{2\nu},\nonumber\\
	y_{-\nu}(x)&=\sum_{n=0}^\infty\frac{(-\nu+1/2)_n}{(-\nu+1)_n(-2\nu+1)_nn!}(-x^2)^nx^{-2\nu},\nonumber\\
	y_0(x)&=\sum_{n=0}^\infty\frac{(1/2)_n}{(\nu+1)_n(-\nu+1)_nn!}(-x^2)^n,
\end{align}
where $(x)_n\equiv\Gamma(x+n)/\Gamma(x)$.
The constants $C^2$ for these solutions are calculated as 
\begin{equation}
	C^2(y_\nu)=C^2(y_{-\nu})=0,\ \ C^2(y_0)=4\nu^2,
\end{equation}
In order to find the solution with the form $\chi=|\chi|e^{i\theta}$ where the amplitude $|\chi|$ is not oscillatory, we have to 
search for $y$ whose constant $C^2$ is negative.
In this case, we know that such solutions are the Hankel functions,
\begin{align}
	H_{\nu}^{(1)}(x)&=J_\nu(x)+iY_\nu(x)=\frac{1}{i \sin\pi\nu}[J_{-\nu}(x)-J_{\nu}(x)e^{-i\pi \nu}],\nonumber\\
	H_{\nu}^{(2)}(x)&=J_\nu(x)-iY_\nu(x)=\frac{1}{i \sin\pi\nu}[J_{\nu}(x)e^{i\pi \nu}-J_{-\nu}(x)],
\end{align}
where 
\begin{align}
	J_\nu(x)&=\sum_{n=0}^\infty\frac{(-1)^n}{n!\Gamma(n+\nu+1)}\left(\frac{x}{2}\right)^{2n+\nu},\nonumber\\
	Y_\nu(x)&=\frac{1}{\sin\pi\nu}[J_\nu(x)\cos\pi\nu-J_{-\nu}(x)],
\end{align}
are the Bessel's functions of the 1st and 2nd kind, respectively.
Comparing the power of the series of solutions, we find that the square of the amplitude $|H^{(1)}_\nu(x)|^2=H^{(1)}_\nu(x) H^{(2)}_\nu(x)$
is written in terms of $y_0$ and $y_{\pm\nu}$ as
\begin{equation}
	H_{\nu}^{(1)}(x)H_{\nu}^{(2)}(x)=\frac{2^{2\nu}\Gamma(\nu)^2}{\pi^2}y_{-\nu}(x)-\frac{2\cos\pi\nu}{\pi\nu\sin\pi\nu}y_0(x)+
	\frac{\Gamma(-\nu)^2}{2^{2\nu}\pi^2}y_\nu(x).
\end{equation}
This result is listed in ref.\cite{HTF1}.
The constant $C^2$ for this solution $y=H_{\nu}^{(1)}H_{\nu}^{(2)}$ turns out to be
\begin{equation}
	C^2=-\frac{16}{\pi^2},
\end{equation}
as desired.
Especially, when $\nu$ is a half-integer ($\nu=1/2+N, N$:integer), we have a polynomial solution,
\begin{equation}
	H_{N+1/2}^{(1)}(x)H_{N+1/2}^{(2)}(x)=\frac{\Gamma(N+1/2)^2}{\pi^2}2^{2N+1}\sum_{n=0}^N\frac{(-N)_n}{(1/2-N)_n(-2N)_nn!}(-x^2)^nx^{-1-2N+2n}.\nonumber
\end{equation}
In this case, we have used analytic continuation since we know the Bessel functions very well. However, if we find the polynomial solution for the third order equation, we can construct the Hankel function from (\ref{eq:solutions}). 
In other words, (\ref{eq:solutions}) represents a new way of writing Hankel functions. 
Figure \ref{fig:bessel} shows $N=1$ case.

These examples tell us that non-oscillatory solutions of the third order equation can be regarded as the evolution of the square of the amplitudes.
We will apply the construction of the solution to the gravitational wave equation in the next section.

\section{Solutions for matter-cosmological constant dominant universe}
 In the previous section, we have shown that the non-oscillatory solution of the third order equation can be regarded as the square of the amplitudes. 
Before applying the formalism to the evolution equation for gravitational waves, we first show that physical observable quantities are directly related
to the solution of the third order equation.
 
The energy density of the relic gravitational wave can be obtained by linear perturbation of the Ricci tensor in flat space as
\begin{equation}
	\rho_{\rm{gw}}=\frac{1}{32\pi G}<\dot{h}_{ij}(t,x)\dot{h}_{ij}(t,x)>=\frac{1}{32\pi G}\int\frac{d^3k}{(2\pi)^3} P(k)|\dot{\chi}(t,k)|^2, \label{energydensity}
\end{equation}
where $P(k)$ is the primordial power spectrum and the overdot represents the derivative with respect to $t$, $\dot{\chi}\equiv d\chi/dt$.
The spectrum of the gravitational waves is usually described by the fraction of the energy density per logarithic frequency interval,
\begin{equation}
	\Omega_{\rm{gw}}=\frac{1}{\rho_{\rm{crit}}(t)}\frac{d\rho_{\rm{gw}}}{d\ln k},
\end{equation}
where $\rho_{\rm{crit}}=3H^2/8\pi G$ is the critical energy density of the universe. Substituting equation (\ref{energydensity}) and expressing it in the conformal time, we get
\begin{equation}
	\Omega_{\rm{gw}}=\frac{k^3}{24\pi^2 a^2(\eta) H^2(\eta)}P(k)\left|{\chi^\prime(\eta,k)}\right|^2.
\end{equation}
If we set $y=|\chi(\eta,k)|^2$, then by using equation (\ref{eq:primeprime}) we can express $\Omega_{\rm{gw}}$ in terms of $y$,
\begin{equation}
	\Omega_{\rm{gw}}=\frac{k^3}{24\pi^2 a^2(\eta) H^2(\eta)}P(k)\left(\frac{1}{2}y^{\prime\prime}+\frac{a^\prime}{a}y^\prime+k^2y\right). \label{intensityiny}
\end{equation}
Although we have to take time average for an oscillatory solution $y$, 
such a procedure is not necessary if we can find a non-oscillatory solution, e.g. polynomial solution, for the third order equation (\ref{eq:thirdorder}) directly.
Energy and momentum of gravitational waves are expressed by the quadratic of the amplitude, therefore the direct analysis of the quadratic is also related to the analysis of these physical quantities.

\begin{figure}[t]
\centering
\includegraphics[clip,width=10.0cm]{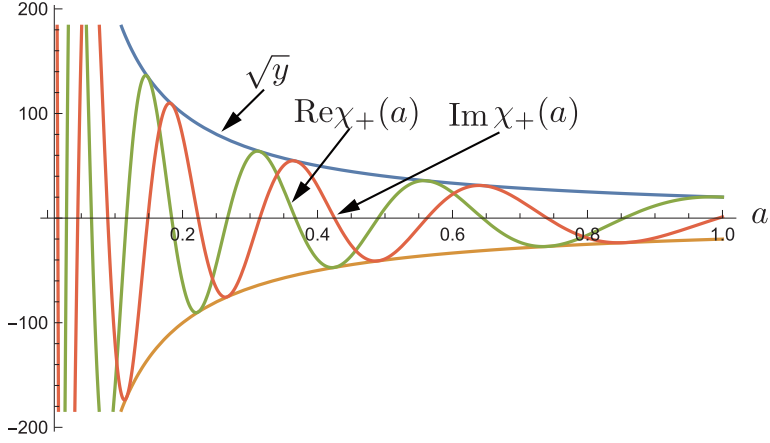}
\caption{Real and imaginary part of the solution $\chi_{+}(a)$ for $k/H_{0}=10, \Omega_m=0.3$, and $\Omega_\Lambda=0.7$.}
\label{fig:solutions1}
\end{figure}

Now, let us apply the construction of the solution to the equation (\ref{eqchiina}). The corresponding third order equation (\ref{eq:thirdorder}) becomes 
\begin{equation}
	\left[\sum_{n=0}^4b_na^n\theta(\theta+n+2)\left(\theta+1+\frac{n}{2}\right)+\frac{4k^2a^2}{H_0^2}(\theta+2)\right]y=0, \label{thirdeq}
\end{equation}
where
\begin{equation}
	\theta=a\frac{d}{da}, \  b_0=\Omega_r,  \ b_1=\Omega_m, \ b_2=\Omega_K, \ b_3=0, \ b_4=\Omega_\Lambda.
\end{equation}
In general we can get power series solutions. 
However, in the case where the radiation is negligible, namely $b_0=0$, equation (\ref{thirdeq}) allows the following  polynomial solution,
\begin{equation}
	y=\frac{b_1}{a^3}+(-3b_2+4(k/H_0)^2)\frac{1}{a^2}-\frac{b_4(3b_2-4(k/H_0)^2)}{(k/H_0)^2}. \label{ysol}
\end{equation}
We can calculate $C^2$ for this solution by using (\ref{eq:csquare}) as follows,
\begin{equation}
	C^2=\frac{b_2-(k/H_0)^2}{(k/H_0)^2}[27b_1^2b_4+4(k/H_0)^2(3b_2-4(k/H_0)^2)^2].
\end{equation}
Furthermore, if we set $b_2=0$, one can see that $C^2$ is negative for any $k\in\mathbb{R}$.
In that case, the solution (\ref{eq:solutions}) for the second order equation becomes
\begin{equation}
	\chi_{\pm}=\sqrt{y}\exp\left(\pm\frac{i{\tilde{C}}}{2}\int \frac{da}{g^{\frac{1}{2}}a^2y(a)}\right), \label{chisol}
\end{equation}
where
\begin{align}
	y=\frac{b_1}{a^3}+\frac{4(k/H_0)^2}{a^2}+4b_4, \label{squareamp}\\
\tilde{C}=(27b_1^2b_4+64(k/H_0)^6)^{\frac{1}{2}}.
\end{align}
Note that the time ($a$) average of the power is proportional to $y$, which depends on the wavelength.
$\chi_+$ and $\sqrt{y}$ are shown in figure \ref{fig:solutions1}.

\section{The Explicit solutions}
\begin{figure}[t]
\centering
\includegraphics[clip,width=12cm]{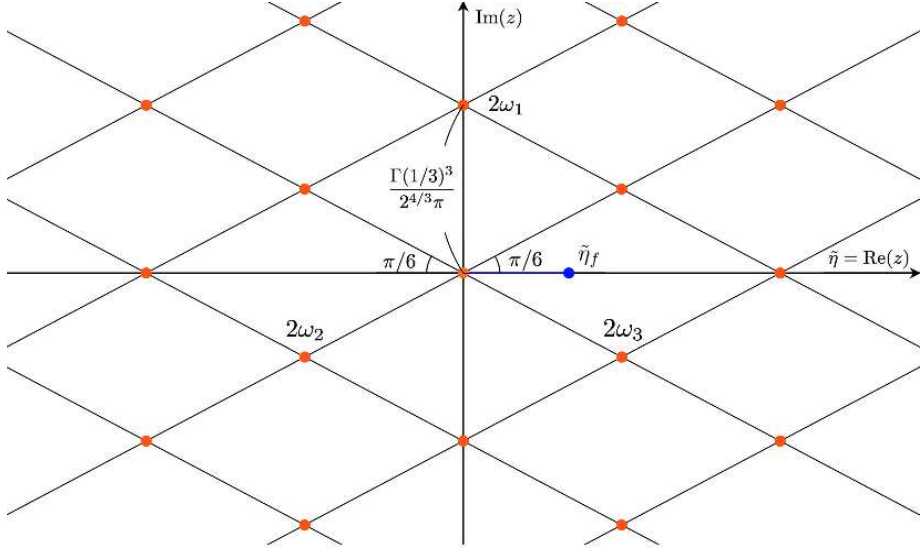}
\caption{Fundamental periods of $\wp(z)$ on the complex plane. Expanding universe corresponds to the line segment connecting the origin and $z=\tilde{\eta}_f$
 (blue point) on the real axis. \label{period}}
\end{figure}

We have constructed gravitational wave solutions in the epoch where radiation and curvature components are negligible
\footnote{Note that although $y$ cannot be regarded as the square of the amplitude in the case of positive $b_2$ because $y$ could be negative, 
the expression (\ref{chisol}) for the solution is still valid for $b_2\neq0$.}.
In this section, we will show that the solutions (\ref{chisol}) can be expressed by elliptic functions if we use the conformal time instead of the scale factor.
We first rescale the scale factor $a$ in such a way that $a=1$ when the energy density of the  matter and that of the cosmological constant component are equal.
We call the Hubble constant at that time $H_{\rm{eq}}$. Then the Friedmann equation (\ref{eq;FE2}) reduces to
\begin{equation}
	\frac{da}{d\eta}=\frac{H_{\rm{eq}}}{\sqrt{2}}(a+a^4)^{\frac{1}{2}}. \label{FE3}
\end{equation}
By using the reciprocal of the scale factor 
\begin{equation}
	x=\frac{1}{a},
\end{equation}
integration of (\ref{FE3}) gives
\begin{equation}
	\frac{1}{2}\int_{1/a}^\infty \frac{dx}{(1+x^3)^{\frac{1}{2}}}=\frac{H_{\rm{eq}}}{2\sqrt{2}}\eta, \label{inva}
\end{equation}
where we have set the integration constant so that $a=0$ at $\eta=0$.
We can solve this relation for $a$ by introducing the Weierstrass's elliptic function $\wp(z)$ (see, for example, \cite{HTF2}),
\begin{equation}
	\wp(z)\equiv\frac{1}{z^2}+\sum_{m,n}{}^{'}\left[\frac{1}{(z-\omega_{m,n})^2}-\frac{1}{\omega_{m,n}^2}\right]. \label{wpdef}
\end{equation}
where $\omega_{m,n}=2m\omega_1+2n\omega_2$ and 
$\sum_{m,n}^{'}$ sums over all integers $m,n$ except for $(m,n)=(0,0)$.
$\omega_1$ and $\omega_2$ are half periods, meaning $\wp(z+\omega_{m,n})=\wp(z)$ for any $z\in\mathbb{C}$ and $m,n\in\mathbb{Z}$.
The inverse of $\wp(z)$ is known as
\begin{equation}
	z=\int_{\wp(z)}^\infty\frac{du}{(4u^3-g_2u-g_3)^{1/2}}, \label{invwp}
\end{equation}
where $g_2$ and $g_3$ are constants determined by the half periods,
\begin{equation}
	g_2=60\sum_{m,n}{}^{'}\frac{1}{\omega_{m,n}^4},\ \ g_3=140\sum_{m,n}{}^{'}\frac{1}{\omega_{m,n}^6}.
\end{equation}
Comparing the equation (\ref{invwp}) and (\ref{inva}), we can see the scale factor is written in terms of the elliptic function with $g_2=0, g_3=-4$,
\begin{equation}
	a(\eta)=\frac{1}{\wp(\tilde{\eta})},\ \ \tilde{\eta}=\frac{H_{\rm{eq}}}{2\sqrt{2}}\eta. \label{asol}
\end{equation}
The half periods are obtained by the formula 
\begin{equation}
	\omega_j=\frac{1}{2}\int_{-\infty}^{e_j}\frac{dx}{(x^3+1)^{\frac{1}{2}}}=-e_ji\frac{2^{-7/3}\Gamma(1/3)^3}{\pi}, \ (j=1,2,3)
\end{equation}
where $e_j=-e^{2\pi i(j-1)/3}\ (j=1,2,3)$ are the roots of $x^3+1=0$. The fundamental periods are shown in figure \ref{period}.
We can see that $a=\infty$ corresponds to a zero of $\wp(z)$, where the conformal time $\tilde{\eta}$ takes the following value,
\begin{equation} 
	\tilde{\eta}_f=\frac{1}{2}\int_0^\infty\frac{dx}{(1+x^3)^{1/2}}=\frac{\Gamma(1/3)^3}{2^{4/3}\sqrt{3}\pi},
\end{equation}
which is exactly $2/3$ of the real half periods
\begin{equation}
	\tilde{\eta}_f=\frac{2}{3}(\omega_3-\omega_2).
\end{equation}

Before considering the gravitational waves, let us briefly see the behavior of this scale factor in the matter-dominant 
and $\Lambda$-dominant eras. More detailed accounts are given in Appendix \ref{limits}.
Since $\wp(z)\sim1/z^2$ around $z\sim0$, scale factor behaves as $a(\eta)\sim\tilde{\eta}^2$ in the matter-dominant era ($\tilde{\eta}\ll1$).
In order to see the behavior in the $\Lambda$-dominant era ($\tilde{\eta}\sim\tilde{\eta}_f$), we use
the following differential equation for $\wp(z)$:
\begin{equation}
	(\wp'(z))^2=4(\wp^3(z)+1). \label{diffeqwp}
\end{equation}
Then we can expand $\wp(z)$ around $\tilde{\eta}=\tilde{\eta}_f$ as 
\begin{equation}
	\wp(\tilde{\eta})=2(\tilde{\eta}_f-\tilde{\eta})+O((\tilde{\eta}_f-\tilde{\eta})^4),
\end{equation}
which leads to the de-Sitter expansion
\begin{equation}
	a(\eta)\sim\frac{1}{2(\tilde{\eta}_f-\tilde{\eta})}.
\end{equation}

It is also instructive to get the relation between the cosmic time $t$ and the conformal time $\eta$. Since 
the Friedmann equation can be solved as
\begin{equation}
	a(t)=\left[\sinh\left(\frac{3}{2\sqrt{2}}H_{\rm{eq}}t\right)\right]^{2/3},
\end{equation}
we can get the following relation,
\begin{equation}
	t=\frac{2\sqrt{2}}{3H_{\rm{eq}}}\left[-\frac{3}{2}\ln \wp(\tilde{\eta})+\ln\left(1+\sqrt{1+ \wp^3(\tilde{\eta})}\right)\right].
\end{equation}

Let us write the solutions (\ref{chisol}) in terms of the conformal time and perform the integration in the phase factor explicitly.
$y$ is written as
\begin{equation}
	y=\frac{1}{2}\wp^3(\tilde{\eta})+4\tilde{k}^2\wp^2(\tilde{\eta})+2=\frac{1}{2}(\wp(\tilde{\eta})-\wp(c_1))(\wp(\tilde{\eta})-\wp(c_2))(\wp(\tilde{\eta})-\wp(c_3)), 
	\label{soly}
\end{equation}
where $\wp(c_j)\ (j=1,2,3)$ are roots of the third order equation $x^3+8\tilde{k}^2x^2+4=0$ and we introduced $\tilde{k}=k/H_{\mathrm{eq}}$.
Then, the integral in the exponent of (\ref{chisol}) is decomposed as
\begin{align}
	\int\frac{da}{g^{\frac{1}{2}}a^2y}&=\int\frac{4\sqrt{2}\wp^2(\tilde{\eta})d\tilde{\eta}}{(\wp(\tilde{\eta})-\wp(c_1))(\wp(\tilde{\eta})-\wp(c_2))(\wp(\tilde{\eta})-\wp(c_3))}\nonumber\\
	&=\frac{4\sqrt{2}\wp^2(c_1)}{(\wp(c_1)-\wp(c_2))(\wp(c_1)-\wp(c_3))}\int\frac{d\tilde{\eta}}{\wp(\tilde{\eta})-\wp(c_1)}
		+(\text{cyclic : } 1\to2\to3\to1).
\end{align}
In order to perform the integral, we introduce the following functions\cite{HTF2}:
\begin{align}
	\zeta(z)&=\frac{1}{z}+\sum_{m,n}{}^{'}\left[\frac{1}{z-\omega_{m,n}}+\frac{1}{\omega_{m,n}}+\frac{z}{\omega_{m,n}^2}\right],\nonumber\\
	\sigma(z)&=z\prod_{m,n}{}^{'}\left[\left(1-\frac{z}{\omega_{m,n}}\right)\exp\left(\frac{z}{\omega_{m,n}}+\frac{z^2}{2\omega_{m,n}^2}\right)\right],
\end{align}
where product $\prod_{m,n}^{'}$ is taken over all integers $m, n$ except for $(m,n)=(0,0)$.
These functions are related to $\wp$ as
\begin{equation}
	\wp(z)=-\zeta^\prime(z),\qquad \zeta(z)=\frac{\sigma^\prime(z)}{\sigma(z)}. \label{sigmazeta}
\end{equation}
There are some addition theorems for these functions, among which we use the following formula:
\begin{equation}
	\zeta(u+v)=\zeta(u)+\zeta(v)+\frac{1}{2}\frac{\wp^\prime(u)-\wp^\prime(v)}{\wp(u)-\wp(v)}. \label{zetaformula}
\end{equation}
Using this formula, we find
\begin{equation}
	\int\frac{d\tilde{\eta}}{\wp(\tilde{\eta})-\wp(c_i)}=\frac{2}{\wp'(c_i)}\left[\zeta(c_i)\tilde{\eta}-\ln\left(\frac{\sigma(\tilde{\eta}+c_i)}{\sigma(\tilde{\eta})
	\sigma(c_i)}(\wp(\tilde{\eta})-\wp(c_i))^{-1/2}\right)\right].
\end{equation}
Then, we obtain the following expression of the solution (\ref{chisol}),
\begin{equation}
	\chi_{\pm}=\frac{1}{\sqrt{2}}\prod_{j=1}^{3}(\wp(\tilde{\eta})-\wp(c_j))^{\frac{1}{2}}
	\biggl[\frac{\sigma(\tilde{\eta}+c_j)}{\sigma(\tilde{\eta})\sigma(c_j)}e^{-\zeta(c_j)\tilde{\eta}}(\wp(\tilde{\eta})-\wp(c_j))^{-1/2}\biggr]^{\mp 4\sqrt{2}i {\tilde{C}}\Theta_j}, \label{explicitsol}
\end{equation}
where
\begin{align}
	{\tilde{C}}&=\sqrt{\frac{27}{8}+64\tilde{k}^6}, \\
	\Theta_1&=\frac{\wp^2(c_1)}{\wp'(c_1)(\wp(c_1)-\wp(c_2))(\wp(c_1)-\wp(c_3))},
\end{align}
and $\Theta_{2,3}$ are given by the cyclic permutation of $c_j$.
For completeness, we note that $\wp(c_j)$ and $\wp'(c_j) \ (j=1,2,3)$ are explicitly given by
\begin{align}
	\wp(c_j)&=-\frac{8\tilde{k}^2}{3}+\frac{32\tilde{k}^4}{3}\frac{e_j}{(\tilde{C}-\sqrt{27/8})^{2/3}}+
		\frac{2}{3}e^*_j\left(\tilde{C}-\sqrt{\frac{27}{8}}\right)^{2/3}, \label{wpcn}\\
	(\wp'(c_j))^2&=4(\wp^3(c_j)+1).
\end{align}
We can further simplify the solution (\ref{explicitsol}) by using another formula
\begin{equation*}
	\wp(v)-\wp(u)=\frac{\sigma(u+v)\sigma(u-v)}{\sigma^2(u)\sigma^2(v)},
\end{equation*}
inside the bracket,
\begin{equation}
	\chi_\pm=\frac{1}{\sqrt{2}}\prod_{j=1}^3(\wp(\tilde{\eta})-\wp(c_j))^{1/2}\left[\frac{\sigma(c_j-\tilde{\eta})}{\sigma(c_j+\tilde{\eta})}
		e^{2\tilde{\eta}\zeta(c_j)}\right]^{\pm2\sqrt{2} i\tilde{C}\Theta_j}.\label{finalsol}
\end{equation}

Although this expression seems complicated, it turns out that this solution includes, as limiting cases, the solutions in matter-dominant and
cosmological constant-dominant era. Since our convention for the scale factor $a$ is such that $a=1$ at matter-$\Lambda$ equality, we have to 
introduce another scale factor $A$ taking arbitrary value $A=A_{\mathrm{eq}}$ at the equality which depends on the energy fraction
$\Omega_m$ and $\Omega_\Lambda$ at $A=1$ in taking such limits. These two scale factors are related as $a=A/A_{\mathrm{eq}}$.
With this re-definition of the scale factor, conformal time $\eta$ and comoving momentum $k$ are also transformed into new ones as
\begin{equation*}
	\tau=A_{\mathrm{eq}}\eta+\mathrm{const.},\ \ q=A_{\mathrm{eq}}k.
\end{equation*}
Appendix \ref{limits} is devoted to detailed calculation of the limits $\Omega_m\to0$ and $\Omega_\Lambda\to0$ showing
\begin{align*}
	\frac{\chi_\pm(\tilde{\eta})}{\chi_\pm(\tilde{\eta}_f)}&\to e^{\pm iq\tau}(1\mp iq\tau), \ (\Omega_m\to0), \\
	\chi_\pm(\tilde{\eta})&\to\frac{1}{\sqrt{2\Omega_\Lambda}}\left(\frac{2q}{H_0}\right)^3\frac{e^{\pm iq\tau}}{(q\tau)^3}(1\mp iq\tau),\ (\Omega_\Lambda\to0).
\end{align*}
These are the well-known solutions expressed by the Hankel functions up to normalization constants.

Also, in the sub-horizon ($\tilde{k}\to\infty$) and super-horizon ($\tilde{k}\to0$) regime, the solution (\ref{finalsol}) reduces to simple forms
as shown in Appendix \ref{klimit}. Equation (\ref{subhor}) shows that leading order in the sub-horizon limit is
\begin{equation*}
	\chi_\pm(\tilde{\eta})\to\mp\frac{2i\tilde{k}}{a(\tilde{\eta})}e^{\pm2\sqrt{2}i\tilde{k}\tilde{\eta}},\ (\tilde{k}\to\infty).
	\label{subhorizon}
\end{equation*}
This is a manifestation that gravitational waves in this regime can be regarded as a collection of massless particles, namely gravitons, because
it follows that the energy density $\rho_{\mathrm{gw}}$ scales as $\rho_{\mathrm{gw}}\propto a^{-4}$ in this regime\cite{GW2}.
To see the super-horizon behavior, we expand the solution (\ref{finalsol}) as $\chi_\pm(\tilde{\eta})=f_\pm(\tilde{\eta})+\tilde{k}^2g_\pm(\tilde{\eta})
+O(\tilde{k}^4)$, where $\tilde{k}=k/H_{\rm{eq}}$. Then, $f_\pm$ and $g_\pm$ are determined in Appendix \ref{klimit} as
\begin{equation}
	\chi_\pm(\tilde{\eta})=-\frac{1}{2\sqrt{2}}(\wp'(\tilde{\eta})\mp2\sqrt{3}i)+\sqrt{2}\tilde{k}^2\left[2\zeta(\tilde{\eta})\mp i
		\sqrt{1+\wp^3(\tilde{\eta})}\int_{\wp(\tilde{\eta})}^\infty\frac{\sqrt{3}xdx}{(1+x^3)^{3/2}}\right]+O(\tilde{k}^4). \label{kexp}
\end{equation}
Considering long wavelength mode that enters the Hubble horizon after the energy density of radiation becomes negligible, 
two independent solutions $\chi_\pm$ should be superposed to form the almost-constant mode $\chi_C$ which behaves $\chi_C\sim$const. at 
the begining of matter-dominant era ($\tilde{\eta}\sim0$).
From (\ref{chiorigin}), we can see the imaginary part of $\chi_\pm$ is such a superposition. Therefore, we set $\chi_C=\mathrm{Im}\chi_+
=(\chi_+-\chi_-)/2i$. By using the $\tilde{k}$ expansion of $\chi_\pm$ (\ref{kexp}), one can read super-horizon behavior of $\chi_C$ as
\begin{equation}
	\chi_C(\tilde{\eta})=\sqrt{\frac{3}{2}}\left[1-2\tilde{k}^2\sqrt{1+\wp^3(\tilde{\eta})}\int_{\wp(\tilde{\eta})}^\infty\frac{xdx}{(1+x^3)^{3/2}}+O(\tilde{k}^4)\right]. \label{chickexp}
\end{equation}
Note that because comoving Hubble parameter $\mathcal{H}=a'/a$ has a minimum $\mathcal{H}_{\mathrm{min}}=\sqrt{3}H_{\mathrm{eq}}/2^{5/6}$ at $a=1/2^{1/3}$, 
modes with low comoving wavenumber $\tilde{k}<\sqrt{3}/2^{5/6}\sim0.97$ never enter the Hubble horizon. 
Therefore expansion (\ref{chickexp}), which is valid for $\tilde{k}\ll1$, is not applicable to modes that enter the horizon in matter-dominant era.
For such modes, we can use the expression (\ref{chisol}) to get the following expression:
\begin{equation}
	\frac{\chi_C(\tilde{\eta})}{\chi_C(0)}=\frac{3}{2\sqrt{2}\tilde{C}}\sqrt{\frac{1}{a^3(\tilde{\eta})}+\frac{8\tilde{k}^2}{a^2(\tilde{\eta})}+4}
		\sin\left[\int_0^{a(\tilde{\eta})}\frac{\sqrt{2}\tilde{C}ada}{\sqrt{a+a^4}(1+8\tilde{k}^2a+4a^3)}\right] \label{chiC}
\end{equation}

\section{Conclusions and Discussions}
We have shown that gravitational wave equation can be solved exactly in the presence of matters and the cosmological constant. 
While the gravitational wave amplitude can be written in terms of elliptic functions, 
square of the amplitude, which is directly related to the energy density,  can be written as a simple polynomial of $1/a$ (\ref{squareamp}).
Because this solution can describe the propagation of gravitational waves of arbitrary wavelength 
in the epoch after energy density of radiation becomes ignorable, which includes the present, it could be used to precise calculations of stochastic
backgrounds of gravitational waves, which are expected to be directly observed in the future, as well as long wavelength modes affecting 
anisotropy of the cosmic microwave background. 
In the former case, a crucial quantity is the intensity (\ref{intensityiny}). Since our solutions are valid after entering matter-dominant era, 
once a matching condition at radiation to matter transition is given, we can complute $\Omega_{\mathrm{gw}}$ exactly.
On the other hand, in the latter case, longer wavelength modes are of main interest. For such a purpose, expression (\ref{chiC}) would be useful.

Unfortunately, we cannot obtain the exact results for the equation in the presence of radiations and matters because the third order equation (\ref{eq:thirdorder}) do not allow any polynomial solutions 
and it is difficult to get non-oscillatory solutions. 
However, it was shown in ref.\cite{TWL} that the transfer function of the amplitudes are obtained numerically as
\begin{equation}
	T(k/k_{\rm{eq}})=[1.0+1.34(k/k_{\rm{eq}})+2.50(k/k_{\rm{eq}})^2]^{1/2},
\end{equation}
where $k_{\rm{eq}}$ is the scale that entered the horizon at matter-radiation equality.
This result indicates that the analysis of the square of the amplitude may be convenient for obtaining the transfer functions analytically.

\appendix
\section{Matter dominant \& $\Lambda$ dominant limit \label{limits}}
Because our consideration treats both the non-relativistic matters and the cosmological constant, all the resuts obtained above should 
include matter-dominant case and $\Lambda$-dominant case. In order to see that, we have to change the scale factor $a$, which is 
normalized such that $a=1$ at the matter-$\Lambda$ equality, to another one $A$ which takes general value $A_{\rm{eq}}$ at the equality.
The relation between these two scale factors is $A=A_{\rm{eq}}a$. Note that this relation induces the relation between the conformal time $\eta$ 
and one associated with $A$, $\tau$, as $\eta=A_{\rm{eq}}\tau+\rm{const.}$.
Introducing the energy fraction of matter $\Omega_m$ and cosmological constant $\Omega_\Lambda$ at $A=1$, the Friedmann equation for $A$ is given as
\begin{equation*}
	\frac{dA}{d\tau}=H_0\sqrt{\Omega_mA+\Omega_\Lambda A^4},
\end{equation*}
where $H_0=(dA/d\tau)_{A=1}$. Comparing this equation with (\ref{FE3}), we get $A_{\rm{eq}}=(\Omega_m/\Omega_\Lambda)^{1/3}$ and
$H_{\rm{eq}}=H_0\sqrt{2/(1+A^3_{\rm{eq}})}$. The matter-dominant case and $\Lambda$-dominant case correspond to the limit $\Omega_\Lambda\to0$ and 
$\Omega_m\to0$, where $A_{\rm{eq}}\to\infty$ and $A_{\rm{eq}}\to0$, respectively. By using the solution (\ref{asol}), $A$ is written as
\begin{equation}
	A(\tau)=A_{\rm{eq}}\left[\wp\left(\frac{H_{\rm{eq}}A_{\rm{eq}}}{2\sqrt{2}}(\tau-\tau_0)\right)\right]^{-1}, \label{scaleA}
\end{equation}
where $\tau_0$ is a constant.
In the following subsections, we will give the behavior of the scale factor and the solution $\chi_\pm$ under the limit of $\Omega_m\to0$ and
$\Omega_\Lambda\to0$.

\subsection{$\Lambda$-dominant case : $\Omega_m\to0$}
Because $\Lambda$-dominant era corresponds to $\tilde{\eta}\sim\tilde{\eta}_f$, we expand $\wp(\tilde{\eta})$ around there.
We set $\tau_0$ so that $\tilde{\eta}=\tilde{\eta}_f$ at $\tau=0$.
We can obtain the explicit form by using the differential equation $\wp'(z)=-2\sqrt{\wp^3(z)+1}$ and the definition of $\tilde{\eta}_f$, i.e. 
$\wp(\tilde{\eta}_f)=0$. The result is 
\begin{equation}
	\wp(\tilde{\eta})=-2(\tilde{\eta}-\tilde{\eta}_f)+2(\tilde{\eta}-\tilde{\eta}_f)^4+O((\tilde{\eta}-\tilde{\eta}_f)^7).
	\label{wpexpansionetaf}
\end{equation}
By substituting this into (\ref{scaleA}), and taking the limit $\Omega_m\to0$, we obtain
\begin{equation}
	A(\tau)\rightarrow-\frac{1}{H_0\tau},
\end{equation}
which represents exactly the de-Sitter expansion.

Before taking the limit of wave function $\chi_\pm$, we expand it around $\tilde{\eta}=\tilde{\eta}_f$.
By using equations (\ref{diffchi}) and (\ref{diffdiffchi}), we get 
\begin{align*}
	\frac{d\chi_\pm}{d\tilde{\eta}}(\tilde{\eta}_f)&=\chi_\pm(\tilde{\eta}_f)\sum_{j=1}^3\frac{1\mp2\sqrt{2}i\tilde{C}\Theta_j\wp'(c_j)}{\wp(c_j)}, \\
	\frac{d^2\chi_\pm}{d\tilde{\eta}^2}(\tilde{\eta}_f)&=\chi_\pm(\tilde{\eta}_f)\left[\sum_{j=1}^3\frac{1\mp2\sqrt{2}i\tilde{C}\Theta_j\wp'(c_j)}{\wp(c_j)}\right]^2 \\
	&\hspace{3em}
		+\frac{\chi_\pm(\tilde{\eta}_f)}{2}\sum_{j=1}^3\left[\frac{\wp''(\tilde{\eta}_f)}{-\wp(c_j)}+\frac{2(-2\pm4\sqrt{2}i\tilde{C}\Theta_j
		\wp'(c_j))}{\wp^2(c_j)}\right].
\end{align*} 
Summations are evaluated by using equations (\ref{sum1}) - (\ref{sum4}), which lead to $d\chi_\pm/d\tilde{\eta}|_{\tilde{\eta}_f}
=0$ and $d^2\chi_\pm/d\tilde{\eta}^2|_{\tilde{\eta}_f}=-8\tilde{k}^2\chi_\pm(\tilde{\eta}_f)$. Thus, 
\begin{align*}
	\chi_\pm(\tilde{\eta})&=\chi_\pm(\tilde{\eta}_f)\left\{1-4\tilde{k}^2(\tilde{\eta}-\tilde{\eta}_f)^2+O((\tilde{\eta}-\tilde{\eta}_f)^3)\right\}\\
	&=\chi_\pm(\tilde{\eta}_f)\left\{1-\frac{1}{2}k^2(\eta-\eta_f)^2+O((\eta-\eta_f)^3)\right\}.
\end{align*}

Finally, we consider the limit $\Omega_m\to0$ of the gravitational wave (\ref{finalsol}). Square of absolute value of the wave function $|\chi_\pm|^2
=y$ is given by (\ref{soly}). We note that the comoving wavenumber $k$ is the physical wavenumber at $a=1$, which is related to the comoving
wavenumber $q$ for the scale factor $A$ as $k=q/A_{\rm{eq}}$. Thus, what should be kept constant in this limit is not $k$ but $q$. Then, $y$ approaches
\begin{equation*}
	2y=\left(\frac{A_{\rm{eq}}}{A}\right)^3+8\left(\frac{q}{A_{\rm{eq}}H_{\rm{eq}}}\right)^2\left(\frac{A_{\rm{eq}}}{A}\right)^2+4
		\to4(1+q^2\tau^2).
\end{equation*}
Next we pick up the phase factor of $\chi_\pm$ normalized by the value at $\tilde{\eta}=\tilde{\eta}_f$,
\begin{equation}
	\prod_{j=1}^3\left[\frac{\sigma(c_j-\tilde{\eta})\sigma(c_j+\tilde{\eta}_f)}{\sigma(c_j+\tilde{\eta})\sigma(c_j-\tilde{\eta}_f)}e^{2\zeta(c_j)(\tilde{\eta}-\tilde{\eta}_f)}\right]^{\pm2\sqrt{2}i\tilde{C}\Theta_j}.
		\label{phasefactor}
\end{equation}
Because $k\to\infty$ as $\Omega_m\to0$, we consider $c_n$ under the limit $k\to\infty$. The limit of $\wp(c_j)$ is easily obtained from the equation
(\ref{wpcn}) as
\begin{align}
	\wp(c_1)&=-8\tilde{k}^2-\frac{1}{16\tilde{k}^4}+O(\tilde{k}^{-10})\to-\infty, \notag\\
	\wp(c_2)&=-\frac{i}{\sqrt{2}\tilde{k}}+\frac{1}{32\tilde{k}^4}+O(\tilde{k}^{-7})\to-i0, \notag\\
	\wp(c_3)&=\frac{i}{\sqrt{2}\tilde{k}}+\frac{1}{32\tilde{k}^4}+O(\tilde{k}^{-7})\to+i0. \label{wpcnexpansion}
\end{align}
Since the elliptic function $\wp(z)$ has second-order poles at $z=2n\omega_1+2m\omega_2 \ ({}^\forall n, m\in\mathbb{Z})$,
$c_1$ approaches one of these poles. However, at the same time, the phase factor (\ref{phasefactor}) is invariant under $c_j\to c_j+2n\omega_1+
2m\omega_2$, therefore we can take $c_1$ such that $c_1\to0$ with $k\to\infty$.
By the similar argument, we can take $c_{2,3}$ such that $c_{2,3}\to\tilde{\eta}_f$. In order to get the expansion of $c_j$ around $k=\infty$, we 
use the expansion of $\wp(\tilde{\eta})$ around $\tilde{\eta}=0$ and $\tilde{\eta}=\tilde{\eta}_f$, the latter of which is given by 
(\ref{wpexpansionetaf}). The former is known as 
\begin{equation}
	\wp(\tilde{\eta})=\frac{1}{\tilde{\eta}^2}-\frac{1}{7}\tilde{\eta}^4+O(\tilde{\eta}^{10}). \label{wpexpansionorigin}
\end{equation}
These expansions of $\wp(\tilde{\eta})$ determines the expansion of $c_j$ as follows:
\begin{align}
	c_1&=\frac{i}{2\sqrt{2}\tilde{k}}\left(1-\frac{27}{7168}\frac{1}{\tilde{k}^6}+O(\tilde{k}^{-12})\right), \notag\\
	c_2&=\tilde{\eta}_f+\frac{i}{2\sqrt{2}\tilde{k}}\left(1+O(\tilde{k}^{-6})\right), \notag\\
	c_3&=\tilde{\eta}_f-\frac{i}{2\sqrt{2}\tilde{k}}\left(1+O(\tilde{k}^{-6})\right).  \label{cnexpansion}
\end{align}
We must convert these $\tilde{k}$ expansion into the $\Omega_m$ expansion. To do so, we need the expansion of $\tilde{k}$ around $\Omega_m=0$, which 
is easily obtained by the definition $\tilde{k}=q/A_{\rm{eq}}H_{\rm{eq}}$ and the constraint $\Omega_\Lambda=1-\Omega_m$,
\begin{equation}
	\tilde{k}=\frac{q}{\sqrt{2}H_0\Omega_m^{1/3}}\left(1+\frac{\Omega_m}{6}+O(\Omega_m^2)\right). \label{tildekexpansion}
\end{equation}
Also, we need the expansion of $\tilde{\eta}$ provided that $\tau$ and $\tilde{\eta}_f$ are kept fixed. The result is
\begin{equation}
	\tilde{\eta}-\tilde{\eta}_f=\frac{H_0\tau}{2}\Omega_m^{1/3}\left(1-\frac{\Omega_m}{6}
		+O(\Omega_m^2)\right). \label{etaexpansion}
\end{equation}
Note that $\tilde{k}(\tilde{\eta}-\tilde{\eta}_f)=q\tau/2\sqrt{2}$ is kept constant under this limit in full order.

We first compute $n=1$ factor in the product (\ref{phasefactor}). 
As $c_1\pm\tilde{\eta}\to\pm\tilde{\eta}_f$, we need the expansion of $\sigma(z)$ around $z=\pm\tilde{\eta}_f$, which is derived from the relation
(\ref{sigmazeta}) and the definition of $\tilde{\eta}_f$, namely $\wp(\tilde{\eta}_f)=0$, as
\begin{equation}
	\sigma(z)=\sigma(\pm\tilde{\eta}_f)\left\{1+\zeta(\pm\tilde{\eta}_f)(z\mp\tilde{\eta}_f)+O((z\mp\tilde{\eta}_f)^2)\right\}.	
\end{equation}
Then, it is straightforward to get 
\begin{equation}
	\frac{\sigma(c_1-\tilde{\eta})\sigma(c_1+\tilde{\eta}_f)}{\sigma(c_1-\tilde{\eta}_f)\sigma(c_1+\tilde{\eta})}=1+O(\Omega_m).
\end{equation}
Remaining factors are $e^{2\zeta(c_1)(\tilde{\eta}-\tilde{\eta}_f)}$ and $\tilde{C}\Theta_1$. The former is expanded by using
$\zeta(z)=1/z+z^5/35+O(z^7)$ as follows:
\begin{equation}
	e^{2\zeta(c_1)(\tilde{\eta}-\tilde{\eta}_f)}=e^{-2iq\tau}(1+O(\Omega_m^2)).
\end{equation}
Equations (\ref{wpcnexpansion}) enable us to expand the latter factor,
\begin{equation}
	2\sqrt{2}i\tilde{C}\Theta_1=-\frac{1}{2}+O(\Omega_m^3).
\end{equation}
Combining the above results, we find
\begin{equation}
	\left[\frac{\sigma(c_1-\tilde{\eta})\sigma(c_1+\tilde{\eta}_f)}{\sigma(c_1-\tilde{\eta}_f)\sigma(c_1+\tilde{\eta})}e^{2\zeta(c_1)
		(\tilde{\eta}-\tilde{\eta}_f)}\right]^{\pm2\sqrt{2}i\tilde{C}\Theta_1}=e^{\pm iq\tau}(1+O(\Omega_m))\to e^{\pm iq\tau}.
\end{equation}
Similar calculation shows 
\begin{align}
	\left[\frac{\sigma(c_2-\tilde{\eta})\sigma(c_2+\tilde{\eta}_f)}{\sigma(c_2-\tilde{\eta}_f)\sigma(c_2+\tilde{\eta})}e^{2\zeta(c_2)
		(\tilde{\eta}-\tilde{\eta}_f)}\right]^{\pm2\sqrt{2}i\tilde{C}\Theta_2}
		&=(1+iq\tau)^{\mp1/2}(1+O(\Omega_m^{1/3})), \\
	\left[\frac{\sigma(c_3-\tilde{\eta})\sigma(c_3+\tilde{\eta}_f)}{\sigma(c_3-\tilde{\eta}_f)\sigma(c_3+\tilde{\eta})}e^{2\zeta(c_3)
		(\tilde{\eta}-\tilde{\eta}_f)}\right]^{\pm2\sqrt{2}i\tilde{C}\Theta_3}
		&=(1-iq\tau)^{\mp1/2}(1+O(\Omega_m^{1/3})).
\end{align}
As a final result, we find that the exact solution (\ref{finalsol}) converges into
\begin{equation}
	\frac{\chi_\pm(\tilde{\eta})}{\chi_\pm(\tilde{\eta}_f)}\to\sqrt{1+q^2\tau^2}e^{\pm iq\tau}(1+iq\tau)^{\mp1/2}(1-iq\tau)^{\pm1/2}
		=e^{\pm iq\tau}(1\mp iq\tau),
\end{equation}
which is exactly the known result for the case of de-Sitter expansion.

\subsection{matter-dominant case : $\Omega_\Lambda\to0$}
In this case, we use the expansion around $\tilde{\eta}=0$ (\ref{wpexpansionorigin}).
Here we set $\tau_0=0$ so that $\tau=0$ corresponds to $\tilde{\eta}=0$. 
Because $\tilde{\eta}=O(\Omega_\Lambda^{1/6})$, the higher order terms in the above expansion doesn't contribute to the limit
$\Omega_\Lambda\to0$. We get
\begin{equation}
	A(\tau)\rightarrow\left(\frac{H_0\tau}{2}\right)^2,
\end{equation}
which is well-known expansion in the matter-dominant era.

As in the previous case, we expand the wave function $\chi_\pm$ around $\tilde{\eta}=0$. 
$\wp(\tilde{\eta})$ is expanded as (\ref{wpexpansionorigin}). $\sigma(z)$ is expanded around $z=z_0$ as
\begin{equation*}
	\log\frac{\sigma(z)}{\sigma(z_0)}=(z-z_0)\zeta(z_0)-\sum_{n=2}^\infty\frac{(z-z_0)^n}{n!}\wp^{(n-2)}(z_0),
\end{equation*}
where we used the relations (\ref{sigmazeta}). Therefore we obtain
\begin{align}
	\sqrt{2}\chi_\pm(\tilde{\eta})&=\prod_{j=1}^3\left(\frac{1}{\tilde{\eta}^2}-\wp(c_j)+O(\tilde{\eta}^2)\right)^{1/2}
		\left(1-\frac{\tilde{\eta}^3}{3!}\wp'(c_j)+O(\tilde{\eta}^5)\right)^{\mp4\sqrt{2}i\tilde{C}\Theta_j} \notag\\
	&=\frac{1}{\tilde{\eta}^3}\left(1-\frac{\tilde{\eta}^2}{2}\sum_{j=1}^3\wp(c_j)\pm\frac{2\sqrt{2}}{3}i\tilde{C}\tilde{\eta}^3
		\sum_{j=1}^3\Theta_j\wp'(c_j)+O(\tilde{\eta}^4)\right)\notag\\
	&=\frac{1}{\tilde{\eta}^3}\left\{1+\left(\frac{2k}{H_{\rm{eq}}}\right)^2\tilde{\eta}^2\pm\frac{2\sqrt{2}}{3}i\tilde{C}\tilde{\eta}^3
		+O(\tilde{\eta}^4)\right\}. \label{chiorigin}
\end{align}

Let us calculate the limit $\Omega_\Lambda\to0$ of the exact solution (\ref{finalsol}).
The procedure is almost the same as in the case of $\Omega_m\to0$ in the previous section.
In this case, the value of the scale factor at the equality, $A=A_{\rm{eq}}$, diverges as $A_{\rm{eq}}=O(\Omega_\Lambda^{-1/3})$ while 
the Habble parameter at that time becomes zero, $H_{\rm{eq}}=O(\Omega_\Lambda^{1/2})$. With this in mind, we find the square of the absolute value of 
the wave function $|\chi_\pm|^2=y$ behaves as
\begin{equation}
	y\to\frac{1}{2\Omega_\Lambda}\left(\frac{2q}{H_0}\right)^6\frac{1}{(q\tau)^6}(1+(q\tau)^2).
\end{equation}
The treatment of the phase factor doesn't so differ from the case of $\Omega_m\to0$, therefore we show only the results:
\begin{align}
	\left[\frac{\sigma(c_1-\tilde{\eta})}{\sigma(c_1+\tilde{\eta})}e^{2\zeta(c_1)\tilde{\eta}}\right]^{\pm2\sqrt{2}i\tilde{C}\Theta_1}
		&\to e^{\pm iq\tau}\left(\frac{1-iq\tau}{1+iq\tau}\right)^{\pm1/2}, \\
	\left[\frac{\sigma(c_2-\tilde{\eta})}{\sigma(c_2+\tilde{\eta})}e^{2\zeta(c_2)\tilde{\eta}}\right]^{\pm2\sqrt{2}i\tilde{C}\Theta_2}
		&\to1, \\
	\left[\frac{\sigma(c_3-\tilde{\eta})}{\sigma(c_3+\tilde{\eta})}e^{2\zeta(c_3)\tilde{\eta}}\right]^{\pm2\sqrt{2}i\tilde{C}\Theta_3}
		&\to1.
\end{align}
Thus, we obatin
\begin{equation}
	\chi_\pm(\tilde{\eta})\to\frac{1}{\sqrt{2\Omega_\Lambda}}\left(\frac{2q}{H_0}\right)^3\frac{e^{\pm iq\tau}}{(q\tau)^3}(1\mp iq\tau),
\end{equation}
which, apart from the normalization, recovers the solution for matter-dominant era.

\section{Wavenumber dependence of the solution \label{klimit}}
In this appendix, we consider small $\tilde{k}$ and large $\tilde{k}$ limits of the exact solution (\ref{finalsol}).
First, let $f_\pm=\lim_{k\to0}\chi_\pm$, then $f_\pm$ should be a solution for the equation (\ref{chieqtilde}) with $\tilde{k}=0$,
\begin{equation*}
	0=\frac{d^2f_{\pm}}{d\tilde{\eta}^2}-\frac{2\wp'(\tilde{\eta})}{\wp(\tilde{\eta})}\frac{df_\pm}{d\tilde{\eta}}.
\end{equation*}
This zero-mode equation can be solved as 
\begin{equation*}
	f_\pm(\tilde{\eta})=A\int\wp^2(\tilde{\eta})d\tilde{\eta}+B,
\end{equation*}
where $A$ and $B$ are integration constants. Integration in the first term is performed by using the fact $\wp''(z)=6\wp^2(z)$,
\begin{equation*}
	f_\pm(\tilde{\eta})=A\wp'(\tilde{\eta})+B, 
\end{equation*}
where we re-defined the constant $A\to 6A$. We fix these constants by considering the behavior around $\tilde{\eta}=0$.
$f_\pm$ is expanded by setting $k=0$ in (\ref{chiorigin}) as
\begin{equation}
	f_\pm(\tilde{\eta})=\frac{1}{\sqrt{2}\tilde{\eta}^3}\pm\sqrt{\frac{3}{2}}i+O(\tilde{\eta}),
\end{equation}
whereas $\wp'(\tilde{\eta})=-2/\tilde{\eta}^3+O(\tilde{\eta}^3)$. From these equations, we conclude that
\begin{equation}
	f_\pm(\tilde{\eta})=-\frac{1}{2\sqrt{2}}(\wp'(\tilde{\eta})\mp2\sqrt{3}i). \label{zeromode}
\end{equation}

In order to see the super-horizon behavior of $\chi_\pm$, we expand as
\begin{equation}
	\chi_\pm(\tilde{\eta})=f_\pm(\tilde{\eta})+\tilde{k}^2g_\pm(\tilde{\eta})+O(\tilde{k}^4), \label{kexpand}
\end{equation}
where $g_\pm$ will be determined below. Expanding (\ref{finalsol}) in terms of $\tilde{k}$ and directly deriving $g_\pm(\tilde{\eta})$ are 
very hard task, therefore we use the differential equation (\ref{chieqtilde}). Inserting (\ref{kexpand}) into (\ref{chieqtilde}), 
the equation for $g_\pm$ reads
\begin{equation}
	0=\frac{d^2g_\pm}{d\tilde{\eta}^2}-\frac{2\wp'(\tilde{\eta})}{\wp(\tilde{\eta})}\frac{dg_\pm}{d\tilde{\eta}}+8f_\pm. \label{geq}
\end{equation}
This equation is easily solved for $dg_\pm/d\tilde{\eta}$ as
\begin{equation*}
	\frac{dg_\pm}{d\tilde{\eta}}(\tilde{\eta})=-8\wp^2(\tilde{\eta})\int\frac{f_\pm(\tilde{\eta})}{\wp^2(\tilde{\eta})}d\tilde{\eta}
		=-2\sqrt{2}\left[\wp(\tilde{\eta})\pm2\sqrt{3}i\wp^2(\tilde{\eta})\int\frac{d\tilde{\eta}}{\wp^2(\tilde{\eta})}\right].
\end{equation*}
Thus, $g_\pm$ is given as
\begin{equation}
	g_\pm(\tilde{\eta})=2\sqrt{2}\zeta(\tilde{\eta})\mp4\sqrt{6}i\int\wp^2(\tilde{\eta})\left(\int\frac{d\tilde{\eta}}{\wp^2(\tilde{\eta})}\right)d\tilde{\eta}.
\end{equation}
Two integration constants implicitly included in this expression reflect the fact that the equation (\ref{geq}) can determine $g_\pm$ only up to 
the zero modes $A\wp'(\tilde{\eta})+B$. These constants can be determined by considering expansion around $\tilde{\eta}=0$ again. The result is 
\begin{equation}
	g_\pm(\tilde{\eta})=2\sqrt{2}\zeta(\tilde{\eta})\mp4\sqrt{6}i\int_0^{\tilde{\eta}}\wp^2(z)\left(\int_0^z\frac{dw}{\wp^2(w)}\right)dz. \label{gsol}
\end{equation}
We can explicitly perform the first integral of the second term by using $\wp''(\tilde{\eta})=6\wp^2(\tilde{\eta})$,
followed by the changes of integration variables,
\begin{align*}
	6\int_0^{\tilde{\eta}}\wp^2(z)\left(\int_0^z\frac{dw}{\wp^2(w)}\right)dz&=\frac{1}{\wp(\tilde{\eta})}+\wp'(\tilde{\eta})
		\int_0^{\tilde{\eta}}\frac{dz}{\wp^2(z)} \\
	&=a(\tilde{\eta})-\sqrt{1+\frac{1}{a^3(\tilde{\eta})}}\int_0^{a(\tilde{\eta})}\frac{da}{\sqrt{1+1/a^3}} \\
	&=a(\tilde{\eta})-\sqrt{1+\frac{1}{a^3(\tilde{\eta})}}\int_{1/a(\tilde{\eta})}^\infty\frac{dx}{x^2\sqrt{1+x^3}} \\
	&=\frac{3}{2}\sqrt{1+\frac{1}{a^3(\tilde{\eta})}}\int_{1/a(\tilde{\eta})}^\infty\frac{xdx}{(1+x^3)^{3/2}}.
\end{align*}
In the second and third equalities, we have changed the integration variables as $a=1/\wp(z)$ and $x=1/a$ respectively, and done partial integration
to obtain the final expression.

Finally, we consider high-frequency, or short-wavelength limit $\tilde{k}\to\infty$ of the solution (\ref{finalsol}).
Because large-$\tilde{k}$ behavior of $\wp(c_n)$ and $c_n$ are already given in (\ref{wpcnexpansion}) and (\ref{cnexpansion}) respectively, we can straightforwardly
get the following expansion:
\begin{equation}
	\chi_\pm(\tilde{\eta})=\mp\frac{2i\tilde{k}}{a(\tilde{\eta})}e^{\pm2\sqrt{2}i\tilde{k}\tilde{\eta}}
		\left(1\pm\frac{i}{2\sqrt{2}\tilde{k}}\zeta(\tilde{\eta})+O(\tilde{k}^{-2})\right) \ \ (\tilde{k}\to\infty). \label{subhor}
\end{equation}
Note that this expansion is not valid around $\tilde{\eta}\sim0$ and $\tilde{\eta}\sim\tilde{\eta}_f$ since we assumed $\wp(\tilde{\eta})\sim O(1)$ 
in expanding $2y=\wp^3(\tilde{\eta})+8\tilde{k}^2\wp^2(\tilde{\eta})+4$ and $\sigma(c_n\pm\tilde{\eta})$.

\section{Confirmation of the solution (\ref{finalsol})}
Here, we confirm that (\ref{finalsol}) really solves the equation (\ref{eq:chi}) with the scale factor $a(\eta)=1/\wp(\tilde{\eta})$.
In this case, the equation (\ref{eq:chi}) reduces to 
\begin{equation}
	0=\frac{d^2\chi}{d\tilde{\eta}^2}-\frac{2\wp'(\tilde{\eta})}{\wp(\tilde{\eta})}\frac{d\chi}{d\tilde{\eta}}+8\tilde{k}^2\chi, \label{chieqtilde}
\end{equation}
where $\tilde{k}=k/H_{\rm{eq}}$. We first calculate the first derivative of (\ref{finalsol}) as follows:
\begin{align}
	\frac{d\chi_\pm}{d\tilde{\eta}}&=\chi_\pm\sum_{j=1}^3\frac{d}{d\tilde{\eta}}\log\left[(\wp(\tilde{\eta})-\wp(c_j))^{1/2}
		\left(\frac{\sigma(c_j-\tilde{\eta})}{\sigma(c_j+\tilde{\eta})}e^{2\tilde{\eta}\zeta(c_j)}\right)^{\pm2\sqrt{2}i\tilde{C}
		\Theta_j}\right] \notag\\
	&=\chi_{\pm}\sum_{j=1}^3\left[\frac{1}{2}\frac{\wp'(\tilde{\eta})}{\wp(\tilde{\eta})-\wp(c_j)}\mp2\sqrt{2}i\tilde{C}\Theta_j\left\{
		\zeta(c_j+\tilde{\eta})+\zeta(c_j-\tilde{\eta})-2\zeta(c_j)\right\}
		\right] \notag\\
	&=\frac{\chi_{\pm}}{2}\sum_{j=1}^3\frac{\wp'(\tilde{\eta})\pm4\sqrt{2}i\tilde{C}\Theta_j\wp'(c_j)}{\wp(\tilde{\eta})-\wp(c_j)}.\label{diffchi}
\end{align}
To get the second and third line, we used the relation $\zeta(z)=d/dz(\log\sigma(z))$ and (\ref{zetaformula}) respectively.
Then, the second derivative becomes
\begin{align}
	\frac{d^2\chi_\pm}{d\tilde{\eta}^2}&=\frac{\chi_\pm}{4}\left(\sum_{j=1}^3\frac{\wp'(\tilde{\eta})\pm4\sqrt{2}i\tilde{C}\Theta_j\wp'(c_j)}{\wp(\tilde{\eta})-\wp(c_j)}\right)^2 \notag\\
	&\hspace{3em}
		+\frac{\chi_\pm}{2}\sum_{j=1}^3\left[\frac{\wp''(\tilde{\eta})}{\wp(\tilde{\eta})-\wp(c_j)}
			-\frac{\wp'(\tilde{\eta})(\wp'(\tilde{\eta})\pm4\sqrt{2}i\tilde{C}\Theta_j\wp'(c_j))}{(\wp(\tilde{\eta})-\wp(c_j))^2}\right].
		\label{diffdiffchi}
\end{align}
The summations in the above expressions are simplified as follows:
\begin{align}
	\sum_{j=1}^3\frac{1}{\wp(\tilde{\eta})-\wp(c_j)}&=\frac{3\wp^2(\tilde{\eta})-2\sum_{j=1}^3\wp(c_j)\wp(\tilde{\eta})+\wp(c_1)\wp(c_2)+\wp(c_2)\wp(c_3)
		+\wp(c_3)\wp(c_1)}{(\wp(\tilde{\eta})-\wp(c_1))(\wp(\tilde{\eta})-\wp(c_2))(\wp(\tilde{\eta})-\wp(c_3))} \notag\\
		&=\frac{3\wp^2(\tilde{\eta})+16\tilde{k}^2\wp(\tilde{\eta})}{2y},\label{sum1}\\
	\sum_{j=1}^3\frac{\Theta_j\wp'(c_j)}{\wp(\tilde{\eta})-\wp(c_j)}&=\frac{\wp^2(c_1)}{(\wp(\tilde{\eta})-\wp(c_1))(\wp(c_1)-\wp(c_2))
		(\wp(c_1)-\wp(c_3))}+(1\to2\to3\to1) \notag\\
		&=-\frac{\wp^2(\tilde{\eta})}{2y}, \label{sum2}\\
	\sum_{j=1}^3\frac{1}{(\wp(\tilde{\eta})-\wp(c_j))^2}&=
		\frac{3\wp^4(\tilde{\eta})+32\tilde{k}^2\wp^3(\tilde{\eta})+128\tilde{k}^4\wp^2(\tilde{\eta})-24\wp(\tilde{\eta})-64\tilde{k}^2}{4y^2}, \label{sum3}\\
	\sum_{j=1}^3\frac{\Theta_j\wp'(c_j)}{(\wp(\tilde{\eta})-\wp(c_j))^2}&=\frac{\wp^4(\tilde{\eta})-8\wp(\tilde{\eta})}{4y^2}.\label{sum4}
\end{align}
Inserting these results into the derivatives of $\chi_\pm$, we obtain,
\begin{align*}
	\frac{1}{\chi_\pm}\frac{d\chi_\pm}{d\tilde{\eta}}&=\frac{1}{4y}\left\{\wp'(\tilde{\eta})(3\wp^2(\tilde{\eta})+16\tilde{k}^2\wp(\tilde{\eta}))\mp4\sqrt{2}i\tilde{C}
		\wp^2(\tilde{\eta})\right\}, \\
	\frac{1}{\chi_\pm}\frac{d^2\chi_\pm}{d\tilde{\eta}^2}&=\frac{1}{16y^2}\left\{\wp'^2(\tilde{\eta})(3\wp^4(\tilde{\eta})+32\tilde{k}^2\wp^3(\tilde{\eta})+48
		\wp(\tilde{\eta})+128\tilde{k}^2)-32\tilde{C}^2\wp^4(\tilde{\eta})\right\} \\
	&\hspace{3em}
		\mp\frac{2\sqrt{2}i\tilde{C}\wp(\tilde{\eta})\wp'(\tilde{\eta})}{y}+\frac{\wp''(\tilde{\eta})}{4y}(3\wp^2(\tilde{\eta})+16\tilde{k}^2\wp(\tilde{\eta})).
\end{align*}
We now calculate the right hand side of the equation (\ref{chieqtilde}) with the aid of (\ref{diffeqwp}),
\begin{align*}
	&\frac{1}{\chi_\pm}\left[\frac{d^2\chi_\pm}{d\tilde{\eta}^2}-\frac{2\wp'(\tilde{\eta})}{\wp(\tilde{\eta})}\frac{d\chi_\pm}{d\tilde{\eta}}
		+8\tilde{k}^2\chi_\pm\right] \\
	&=\frac{1}{16y^2}\left\{4(\wp^3(\tilde{\eta})+1)(3\wp^4(\tilde{\eta})+32\tilde{k}^2\wp^3(\tilde{\eta})+48\wp(\tilde{\eta})+128\tilde{k}^2)-32
		\tilde{C}^2\wp^4(\tilde{\eta})\right\} \\
	&\hspace{3em}
		\mp\frac{2\sqrt{2}i\tilde{C}\wp(\tilde{\eta})\wp'(\tilde{\eta})}{y}+\frac{3\wp^3(\tilde{\eta})}{2y}(3\wp(\tilde{\eta})+16\tilde{k}^2)
		+8\tilde{k}^2 \\
	&\hspace{4em}
		-\frac{2\wp'(\tilde{\eta})}{\wp(\tilde{\eta})}\frac{1}{4y}\left\{\wp'(\tilde{\eta})(3\wp^2(\tilde{\eta})+16\tilde{k}^2\wp(\tilde{\eta}))
		\mp4\sqrt{2}i\tilde{C}\wp^2(\tilde{\eta})\right\} \\
	&=\frac{1}{16y^2}\left\{4(\wp^3(\tilde{\eta})+1)(3\wp^4(\tilde{\eta})+32\tilde{k}^2\wp^3(\tilde{\eta})+48\wp(\tilde{\eta})+128\tilde{k}^2)
		-32\tilde{C}^2\wp^4(\tilde{\eta})\right. \\
	&\hspace{4em}\left.
		-4(\wp^3(\tilde{\eta})+8\tilde{k}^2\wp^2(\tilde{\eta})+4)(3\wp(\tilde{\eta})+16\tilde{k}^2)(\wp^3(\tilde{\eta})+4)
		+32\tilde{k}^2(\wp^3(\tilde{\eta})+8\tilde{k}^2\wp^2(\tilde{\eta})+4)^2\right\} \\
	&=0.
\end{align*}

\end{document}